\documentclass[prl,superscriptaddress,twocolumn,showpacs,floatfix]{revtex4}

\usepackage{graphicx}
\usepackage{dcolumn}
\usepackage{bm}
\usepackage[latin1]{inputenc}
\usepackage{amsmath}
\usepackage{amssymb}
\usepackage{color}

\begin{document}


\title{Coexistence of Magnetism and Superconductivity in the Iron-based Compound Cs$_{0.8}$(FeSe$_{0.98}$)$_2$}

\author{Z.~Shermadini}
\affiliation{Laboratory for Muon Spin Spectroscopy, Paul Scherrer
Institute, CH-5232 Villigen PSI, Switzerland}
%
\author{A.~Krzton-Maziopa}
 \affiliation{Laboratory for Developments and Methods, Paul Scherrer Institute, CH-5232 Villigen PSI, Switzerland}
\author{M.~Bendele}
\affiliation{Laboratory for Muon Spin Spectroscopy, Paul Scherrer
Institute, CH-5232 Villigen PSI, Switzerland}
\affiliation{Physik-Institut der Universit\"at Z\"urich, Winterthurerstrasse 190, CH-8057 Z\"urich, Switzerland}
\author{R.~Khasanov}
 \affiliation{Laboratory for Muon Spin Spectroscopy, Paul Scherrer
Institute, CH-5232 Villigen PSI, Switzerland}
\author{H.~Luetkens}
\affiliation{Laboratory for Muon Spin Spectroscopy, Paul Scherrer
Institute, CH-5232 Villigen PSI, Switzerland}
\author{K.~Conder}
 \affiliation{Laboratory for Developments and Methods, Paul Scherrer Institute, CH-5232 Villigen PSI, Switzerland}
\author{E.~Pomjakushina}
 \affiliation{Laboratory for Developments and Methods, Paul Scherrer Institute, CH-5232 Villigen PSI, Switzerland}
\author{S.~Weyeneth}
\affiliation{Physik-Institut der Universit\"at Z\"urich, Winterthurerstrasse 190, CH-8057 Z\"urich, Switzerland}
\author{V.~Pomjakushin}
 \affiliation{Laboratory for Neutron Scattering, Paul Scherrer Institute, CH-5232 Villigen PSI, Switzerland}
\author{O.~Bossen}
\affiliation{Physik-Institut der Universit\"at Z\"urich, Winterthurerstrasse 190, CH-8057 Z\"urich, Switzerland}
\author{A.~Amato}
\email[corresponding author:~]{alex.amato@psi.ch}
\affiliation{Laboratory for Muon Spin Spectroscopy, Paul Scherrer
Institute, CH-5232 Villigen PSI, Switzerland}

\begin{abstract}
We report on muon-spin rotation and relaxation ($\mu$SR), electrical resistivity,
magnetization and differential scanning calorimetry measurements performed on a
high-quality single crystal of Cs$_{0.8}$(FeSe$_{0.98}$)$_2$. Whereas our
transport and magnetization data confirm the bulk character of the
superconducting state below $T_c = 29.6(2)$~K, the $\mu$SR data indicate that
the system is {\it magnetic} below $T_{\rm N} = 478.5(3)$~K, where a
first-order transition occurs. The first-order character of the magnetic
transition is confirmed by differential scanning calorimetry data. Taken all
together, these data indicate in Cs$_{0.8}$(FeSe$_{0.98}$)$_2$ a microscopic
coexistence between the superconducting phase and a strong magnetic phase. The
observed $T_{\rm N}$ is the highest reported to date for a magnetic
superconductor.

\end{abstract}

\maketitle
\sloppy


The discovery of superconductivity in the Fe-based systems has triggered a
remarkable renewed interest for possible new routes leading to high-$T_c$
superconductivity \cite{1}. As observed in the cuprates, the iron-based
superconductors, exhibit an interplay between magnetism and superconductivity
suggesting the possible occurrence of unconventional superconducting states.
Other common properties are the layered structure and the low carrier density.
A result of the numerous studies on iron-based systems has been the discovery
of a series of superconducting material based on FeAs-layer, as the so-called
`1111' ({\it Re}FeAsO$_{1-x}$F$_x$,\,{\it Re}\,=\,rare-earth), the `111'
(LiFeAs), the `122' {\it A}Fe$_2$As$_2$ ({\it A}\,=\,K,\,Sr,\,Ba) or the
`22426' (Fe$_2$As$_2${\it Ae}$_4${\it M}$_2$O$_6$ where {\it Ae} is an alkaline
earth metal and {\it M} is a transition metal) families (for review see, {\it
e.g.}, Ref.~\onlinecite{Johnston10} and references therein).

Besides FeAs-layer systems, superconductivity has been reported in the related
compound FeSe$_{1-x}$ (`11' or `011' family) \cite{Hsu08}. This system presents
a remarkable increase of $T_c$ under pressure or by a partial substitution on
the chalcogenide site. In addition, recent muon-spin rotation and relaxation
($\mu$SR) and magnetization studies, performed by some of us \cite{2}, have
revealed the occurrence of antiferromagnetism under pressure (above $\sim
0.8$\,GPa) and its coexistence with superconductivity on short length scales in
the full sample volume. Furthermore, both forms of order appear to be
stabilized by pressure, since $T_c$ as well as $T_N$ and the magnetic order
parameter simultaneously increase with increasing pressure. All these results
establish that FeSe-layer systems are themselves remarkable superconductors and
call for the further study of new FeSe-based superconductor families.

Very recently superconductivity at about 30\,K was reported in the FeSe-layer
compound K$_{0.8}$Fe$_2$Se$_2$ (see Ref.~\onlinecite{3}). This compound was
obtained by solid state reaction leading to a potassium intercalation between
FeSe layers. This system is isostructural to the `122' (i.e. tetragonal
ThCr$_2$Si$_2$ type structure -- space group $I4/mmm$). Soon after, some of us
discovered that the related compound Cs$_{0.8}$(FeSe$_{0.98}$)$_2$ exhibits a
similar superconducting transition temperature ($T_c \simeq 27.4$~K)
\cite{anna}. It was shown, in addition, that rather large single crystals of
Cs$_{0.8}$(FeSe$_{0.98}$)$_2$ can be produced, allowing one to hopefully
performed a full study of the microscopic superconducting and/or magnetic
properties.

These recent discoveries allow one to perform a direct comparison
between both families of the chalcogenide-`122' and pnictide-`122'
systems concerning the interplay between magnetism and
superconductivity. In the pnictide-`122' there is an ongoing
debate about the kind of coexistence of magnetism and
superconductivity in the under-doped region of the phase diagram.
In the alkali-metal ion substituted systems, magnetism and
superconductivity seem to compete since they only coexist in a phase
separated manner \cite{park}. On the other side, when doped on the Fe site the
pnictide-`122' systems apparently show a microscopic coexistence  
as evidenced by NMR measurements \cite{laplace}. Anyhow a
competition between the two forms of order in the latter case is
apparent from neutron scattering works since the magnetic order
parameter is partially suppressed below the superconducting transition temperature
\cite{pratt,christianson}. 
Hence, it appears that in the
pnictide-`122' systems static magnetism has to be destroyed by a
control parameter like doping or pressure before superconductivity
can develop its full strength.

Here we report on $\mu$SR, transport and thermodynamic
measurements on Cs$_{0.8}$(FeSe$_{0.98}$)$_2$ which demonstrate
the occurrence of both static magnetism and superconductivity in
this system. Our measurements unambiguously show that both states
microscopically coexist at low temperatures. Surprisingly, we find
magnetic ordering with an unexpectedly high N\'eel temperature
of $\sim478$~K and at the same time a high superconducting $T_c \simeq 30$~K
showing that both order parameters are quite robust and may
coexist without apparent competition.


A 
single crystal of Cs$_{0.8}$(FeSe$_{0.98}$)$_2$ was grown from the melt using
the Bridgman method \cite{anna}. From this large single crystal, different
samples were obtained by cleaving the crystal along the basal plane of the
tetragonal structure. No significant deviation could be observed on the
physical properties of the different samples. They were characterized by powder
x-ray diffraction using a D8 Advance Bruker AXS diffractometer with Cu
$K_{\alpha}$ radiation. The magnetic properties of the crystals were
investigated by a commercial \textit{Quantum Design} 7~T Magnetic Property
Measurement System MPMS-XL SQUID Magnetometer at temperatures ranging from 2~K to
50~K using the Reciprocating Sample Option. The measurements of resistivity
were done using the \textit{Quantum Design} Physical Properties Measurement
System PPMS-9 in a temperature range from 2~K to 300~K. Differential scanning
calorimetry (DSC) experiments were performed with a Netzsch DSC 204F1 system.
Measurements were performed on heating and cooling with a rate of 5-20~K/min
using 20~mg samples encapsulated in standard Al crucibles. An argon stream was
used during the whole experiment as protecting gas. Finally, zero-field and
transverse-field $\mu$SR data were obtained using the GPS and DOLLY instruments
located on the $\pi$M3 and $\pi$E1 beamlines of the Swiss Muon Source (Paul
Scherrer Institute Villigen, Switzerland). Measurements were performed with
static and dynamical helium flow cryostats between 2 and 315~K and with a Janis
closed-cycle refrigerator between 300 and 500~K.

\begin{figure}[t]
\center{\includegraphics[width=0.80\columnwidth]{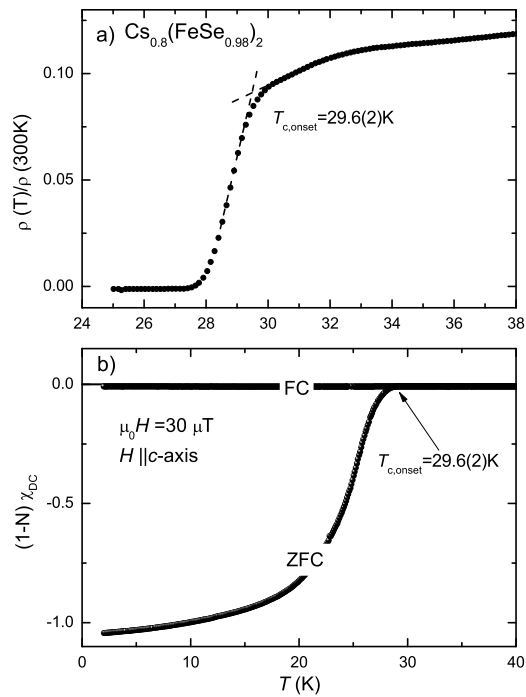}}
    \caption[]{a) Temperature dependence of the electrical resistivity of
    Cs$_{0.8}$(FeSe$_{0.98}$)$_2$ in the vicinity of the superconducting transition.
    b) Temperature dependence of dc magnetic susceptibility 
    for both zero-field cooling (ZFC) and field cooling procedures (FC) obtained
    with a magnetic field     of $\mu_0H=30\,\mu$T applied along the $c$-axis.}
     \label{figure_resis_magnetization}
\end{figure}

The first step of our investigation was to confirm the superconducting ground
state of Cs$_{0.8}$(FeSe$_{0.98}$)$_2$. This was performed by measuring the
electrical resistivity and magnetization. As shown on
Fig,~\ref{figure_resis_magnetization}, these data confirm the occurrence and
bulk character of the superconducting state below $T_c \simeq 29.6$~K. The
resistivity was measured with the electrical current applied in the basal
plane. The magnetization data were obtained in a magnetic field of
$\mu_0H=30\,\mu$T with the crystallographic \textit{c}-axis parallel to the
field.
For the adopted geometry during our magnetization measurements, the
demagnetization factor $N$ of our sample (dimensions:
$2\times3.8\times2$~mm$^3$) was estimated to be 0.44 (see, {\it e.g.},
Ref.~\onlinecite{akishin}) which leads to a superconducting volume fraction
compatible with 100\% of the sample volume.

With this in mind, we turned our focus on the determination of the microscopic
properties by $\mu$SR, performing first zero-field experiments. Generally, the
zero-field $\mu$SR signal for a single crystal can be written as
\cite{yaouanc}:
\begin{equation}
 \label{equation_timeevolpol}
  A(t) = A_0\int f(\mbox{\bf B}_{\mu}){\Big [}\cos^2\theta +
  \sin^2\theta\cos(\gamma_{\mu}B_{\mu}t){\Big ]}d\mbox{\bf B}_{\mu}~,
\end{equation}where $A_0$ is the initial asymmetry, $f(\mbox{\bf B}_{\mu})$
is the magnetic field distribution function at the muon site, $\theta$ is the
angle between the local internal field and the initial muon-spin  polarization
$\mbox{\bf P}_{\mu}(0)$, and $\gamma_{\mu}/(2 \pi)$ is the gyromagnetic ratio
of the muon.
\begin{figure}[t]
\center{\includegraphics[width=0.80\columnwidth]{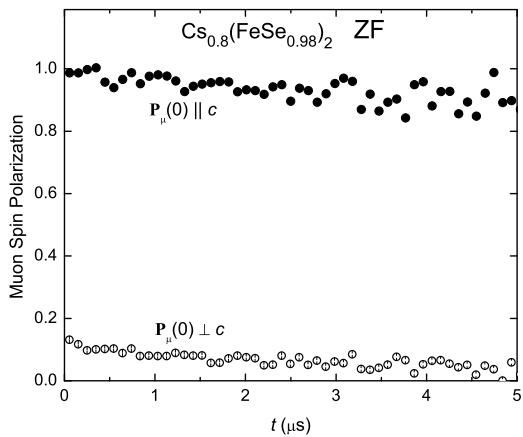}}
    \caption[]{$\mu$SR signals recorded at 2~K with the initial muon
polarization along and perpendicular to the crystallographic $c$-axis. About a
fraction of 0.07 of the signal is due to the cryostat and sample holder signal
which is slowly relaxing. Note the missing amplitude for the geometry
corresponding to the initial muon polarization perpendicular to the $c$-axis.}
\label{figure_zf}
\end{figure}
Surprisingly, as shown on Fig.~\ref {figure_zf}, it appeared that
at low temperatures essentially the full $\mu$SR signal arising
from the sample (more precisely about 95\% of it) is wiped out
when performing zero-field $\mu$SR measurements with
$\mathbf{P}_{\mu}(0)$ oriented along the crystallographic basal
plane. On the other hand, a full and non-depolarizing signal is
obtained when the initial polarization is parallel to the
$c$-axis. Therefore, our observations show that the muon is
sensing {\it spontaneous static internal fields}
$\mathbf{B}_{\mu}$ at low temperatures oriented solely along the
$c$-axis. Such behavior is only observed in long range ordered
magnetic materials with a well defined internal-field direction at the muon
site. It also indicates an homogeneous magnetic state without contributions
of magnetic impurity phases (possessing different magnetic
structures).
In addition, the absence of a detectable $\mu$SR signal when
$\mathbf{P}_{\mu}(0)\perp \mathbf{\hat c}$ [i.e.
$\mathbf{P}_{\mu}(0)\perp \mathbf{B}_{\mu}$] indicates either a
very high value of the internal field (leading to muon spin
precessions much faster than our time resolution) or more probably
a large field distribution along the $c$-axis due to a complicated
magnetic structure with possible large or spatially modulated
ordered moments. Note that neutron scattering measurements
\cite{lumsden} yield an ordered moment of
$\approx$2~$\mu_\textrm{B}$ in the related chalcogenide FeTe in
which a well defined, albeit strongly damped, $\mu$SR precession is observed \cite{rustem}.
This suggests that the ordered magnetic moment in
Cs$_{0.8}$(FeSe$_{0.98}$)$_2$ is at least of the order of
2~$\mu_\textrm{B}$ also.

It should be stressed
that the static, most probably
antiferromagnetic order persists down to 2~K, i.e. well into
the superconducting state. This is especially noteworthy as a part
of the very same crystal which was studied in the magnetization
experiment discussed above is found to exhibit bulk
superconductivity (see Fig.~\ref{figure_resis_magnetization}~b).
Therefore, the $\mu$SR data unambiguously indicate a {\it
microscopic coexistence between magnetism and superconductivity}.

\begin{figure}[t]
\center{\includegraphics[width=0.80\columnwidth]{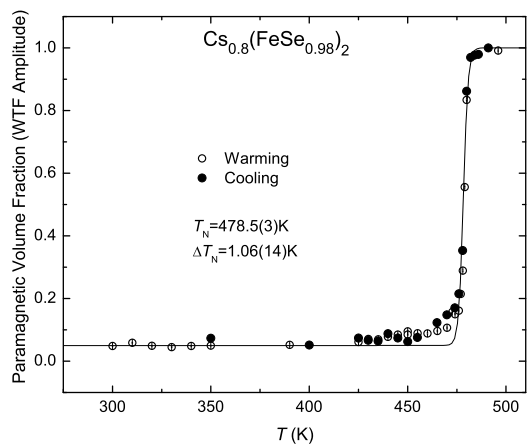}}
    \caption[]{Amplitude of the persisting precession obtained in WTF $\mu$SR experiments ($\mu_0H=3$~mT)
in Cs$_{0.8}$(FeSe$_{0.98}$)$_2$. Such amplitude represents the paramagnetic
volume fraction of the sample as it corresponds to the fraction of muons
stopping in a nonmagnetic environment. The solid line is the fit by using a
Fermi-type function (see text for details). Note the step-like behavior at
$T_{\rm}\simeq479$~K. No clear hysteresis could be observed upon cooling and
warming.} \label{figure_wtf}
\end{figure}

To gain more insight on the magnetic state, we performed zero-field as well as
weak-transverse-field (WTF) $\mu$SR measurements up to 500~K. $\mu$SR
measurements in the WTF configuration are used to determine the volume
fractions  of sample regions with and without static magnetic order. A
persistent oscillation amplitude in WTF $\mu$SR spectra would reflect the
fraction of the muons ensemble (i.e. fraction of the sample volume) with
a nonmagnetic environment. Figure~\ref{figure_wtf} shows the precessing
amplitude of the obtained WTF $\mu$SR signal normalized to its value obtained
in the paramagnetic state. Up to about 475~K, the $\mu$SR amplitude is very
low, indicating that Cs$_{0.8}$(FeSe$_{0.98}$)$_2$ is still in a magnetic
state. Upon increasing the sample temperature above 479~K, a step-like increase
of the WTF $\mu$SR amplitude is observed indicating the transition to a
paramagnetic state at higher temperature \cite{no_diffusion}.
The fact that the whole sample orders at a well defined
ordering temperature again proves the homogeneity of our sample
and excludes possible impurity phases to be present.
The ordering temperature $T_{\rm N}$  was determined by fitting a
Fermi-type function: $\{1+\exp[(T_{\rm
N}-T)/\Delta T_{\rm N}]\}^{-1}$ ($\Delta T_{\rm N}$ is the width of transition) to
the data (solid line in Fig.~\ref{figure_wtf}) \cite{Khasanov08}.
The sharpness of the transition [$\Delta T_{\rm N}=1.06(14)$~K] is compatible
with a first-order transition as confirmed further by our DSC data (see below).
The very high magnetic transition temperature [$T_{\rm N}=478.5(3)$~K, $T_{\rm
N}\simeq 17 \times T_c$] and the possibly high value of the ordered moments, as
also confirmed by the first-principles calculations \cite{Yan10,cao}, indicate a
very robust magnetic state. The observation of a microscopic coexistence
between this strong magnetic state and superconductivity at low temperatures is
rather astonishing and points to an unconventional character for the
superconducting state. Note that the observed $T_{\rm N}$  is, to the best of
our knowledge, the highest reported so far for any kind of magnetic
superconductor.

\begin{figure}[t]
\center{\includegraphics[width=0.80\columnwidth]{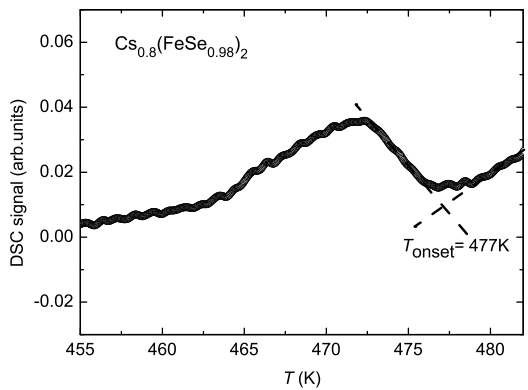}}
    \caption[]{Differential scanning
calorimetry (DSC) signal recorded upon cooling with a sweeping rate of 20~K/min
in Cs$_{0.8}$(FeSe$_{0.98}$)$_2$. The small but clear peak, with an onset
temperature of $\sim$477~K signals an enthalpy of transition $\Delta H$
characterizing a first-order transition. }
 \label{figure_dsc}
\end{figure}

The first-order type of the magnetic transition at $T_{\rm N}=478.5(3)$~K is
confirmed by our DSC measurements reported in Fig.~\ref{figure_dsc}. A small
but definite peak is observed in the data reflecting an enthalpy of transition
due to the first order magnetic transition. The temperature onset is of the
order of 477~K, i.e. perfectly compatible with what observed in the
$\mu$SR data. The small difference might arise from an imperfect sample
thermalization in the rather fast cooled DSC measurements. We observe a slight
temperature hysteresis between the DSC data obtained upon warming (not shown)
and cooling which also seems dependent on the temperature sweep rate used
during the measurements. As DSC measurements have to be performed with a finite
temperature sweeping rate, no real conclusions can be drawn about a possible
temperature hysteresis.
On the other hand, our WTF $\mu$SR data do not show a visible temperature
hysteresis around the transition at $T_{\rm N}\simeq 479$~K.

As said above, the observation of coexistence of magnetism and
superconductivity has already been reported in pnictides-`122'
\cite{park,Sanna09,khasanov_2}. A characteristic of these iron-based family is
that the temperature of the magnetic transition needs to decrease (by doping or
external pressure) prior to observe a superconducting state at low temperature.
Moreover, the ratio between $T_{\rm N}$ and $T_c$ is always much smaller for
samples with $T_c$ near to the optimum than that observed in our present
measurements for Cs$_{0.8}$(FeSe$_{0.98}$)$_2$. In addition $T_c$ is always
found to increase upon decreasing the strength of the magnetic state.
Therefore, it could appear very appealing to try to weaken the magnetic state
in Cs$_{0.8}$(FeSe$_{0.98}$)$_2$ in order to strengthen the superconducting
state (i.e. increase $T_c$). Obviously, additional measurements tracking
the evolution of both transitions as a function of doping and/or pressure are
urgently needed.

On the other hand, the interplay between magnetism and
superconductivity in the chalcogenide iron-based systems might be
rather opposite than the one observed in the pnictides. An 
unusual behavior has been observed in the FeSe$_{1-x}$ family
under pressure \cite{2}. By applying pressures
above 0.8~GPa a static magnetic phase appears which
microscopically coexists with superconductivity. In addition, one observes that 
{\it both} the magnetic \cite{2} and superconducting
\cite{Margadonna09,2} transition temperatures increase with
increasing pressure. In this vein, we note that the pressure
evolution of $T_c$ in FeSe$_{1-x}$ appears first to saturate in
the absence of magnetic state. The subsequent strong increase of
$T_c$, observed upon increasing the pressure above 0.8~GPa, is
concomitant to the occurrence of static magnetism increasing under
pressure. It is therefore plausible to consider that a strong
magnetic state might be even the prerequisite for the observation
of high-$T_c$'s in chalcogenide iron-based systems. 

In summary, we have presented strong evidence that the superconducting state
observed in Cs$_{0.8}$(FeSe$_{0.98}$)$_2$ below 29.6(2)~K is actually
microscopically coexisting with a rather strong magnetic phase with a
transition temperature at 478.5(3)~K. DSC data point to a first-order character
for the magnetic transition, which appears characterized by rather large static
iron-moments as the $\mu$SR signal is wiped out for an initial
muon-polarization perpendicular to the crystallographic $c$-axis.

\acknowledgments Part of this work was performed at the Swiss Muon Source
(S$\mu$S), Paul Scherrer Institute (PSI, Switzerland). The work was partially
supported by the Swiss National Science Foundation. We acknowledge
M.~Mansson for his help during the resistivity measurements. A.K.M.
acknowledges the support by the Scientific Exchange Programme Sciex-NMSch
(Project Code 10.048) and E.P. acknowledges the support by the NCCR MaNEP Project.


\begin{thebibliography}{99}
\bibitem{no_diffusion} Note that we can safely exclude that such transition is arising from muon diffusion in view of its sharpness (no activation energy behaviour observed) and also due to the observation at the same temperature of a transition in the DSC data (see
text).
\end{thebibliography}
\end{document}